# A Web-based Adaptive and Intelligent Tutor by Expert Systems

Hossein Movafegh Ghadirli [1]  and  Maryam Rastgarpour [2]

[1] Graduate student in Computer Engineering, Young Researchers Club, Islamshahr Branch, Islamic Azad University, Islamshahr, Iran. email: hossein.movafegh@iau-saveh.ac.ir

[2] Faculty of Computer Engineering, Department of Computer, Science and Research Branch, Islamic Azad University, Saveh, Iran. email: m.rastgarpour@gmail.com

**Abstract.**    Todays, Intelligent and web-based E-learning is one of regarded topics. So researchers are trying to optimize and expand its application in the field of education. The aim of this paper is developing of E-learning software which is customizable, dynamic, intelligent and adaptive with Pedagogy view for learners in intelligent schools. This system is an integration of adaptive web-based E-learning with expert systems as well. Learning process in this system is as follows. First intelligent tutor determines learning style and characteristics of learner by a questionnaire and then makes his model.  After that the expert system simulator plans a pre-test and then calculates his score. If the learner gets the required score, the concept will be trained. Finally the learner will be evaluated by a post-test. The proposed system can improves the education efficiency highly as well as decreases the costs and problems of an expert tutor.  As a result, every time and everywhere (ETEW) learning would be provided via web in this system. Moreover the learners can enjoy a cheap remote learning even at home in a virtual simulated physical class.  So they can learn thousands courses very simple and fast.

## 1. Introduction

The application of computers in learning began on 1980. Many efforts have been done in order to update and optimize electronic learning (E-learning) yet which great and dramatic advances have been observed in recent years.

Generally, E-learning means to improve educational efficiency using information and communication technology [1].

At the first, some Medias like CDs or web applied for E-learning. But these kinds of education are static, non-intelligent and inflexible. Because the course subject had been organized by prior procedure and then trained to different learner in the same style. Diversity of learners leaded to decrease the efficiency of this style. In fact, repeating some lessons was needed for some learners in this method and also some lessons must be removed for some other learners. Later the researchers in pedagogy sciences (training/learning methods) concluded that the learning must be dynamic and intelligent. The fact is that an expert tutor can adapts the sequence of lessons and speed of training with aptitude and characteristics of learner. He can also adjust the expression style with learner's mood as well as cancels the class due to incorporate mental conditions of the learner.

Nowadays, "web-based learning" and "intelligent learning" is one of the most regarded topics in education [2,3]. Moreover, expert tutor is infrequent and expensive. A web-based tutor has some benefits like tirelessly, predominate on concepts, low cost and invariant of time and place. However millions learners of the world can learn by thousands of expert tutor via web in an intelligent and virtual schools.

This paper introduces an intelligent system to apply the abilities of expert systems. So E-learning would be efficient, adaptive and performed by computer and web. Adaptation of web-based contexts is very important, because the contexts would be used by millions variant learners. So the concept, which is developed for one user, isn't applicable for others [3].

The proposed system determines the learning style by a test. Then the learning process starts. Gradually, some characteristics of learner may be change by learner's progress. These improvements would be saved by system in learning process. So learner model gets more accurate step by step. System can receives scientific and mental feedback of learner expertise and intellectually and then change the learning style during the process. This web-based system is developed to facilitate learning every time and everywhere (ETEW). Web-based content is installed and supported in one place while millions learners of the world can use it just via a computer connected to the internet [4]. The aim of proposed system is that to offer the content which the user is not aware about it.

The rest of this paper is organized as follows. Section 2 defines an intelligent E-learning system and presents some available samples. Then it deliberates adaptive E-learning and some learning styles in section 3. Section 4 describes the proposed E-learning system which is intelligent, adaptive, customized and web-based. Finally this paper concludes in section 5.

## 2. Intelligent E-learning System

The adaptive intelligent systems are not novel at all. All of these systems are a kind of "Intelligent Tutoring Systems (ITS)" or "Adaptive Hypermedia Systems" [5]. E-learning systems are categorized into two classes: intelligent and non-intelligent. Non-intelligent learning is static, inflexible as mentioned. In these systems, tutor develops course topics previously. Then software engineer presents them in variant methods and the same style to learners. The same style for variant learners is the biggest disadvantage of these systems. Since there are different kinds of learners in E-learning systems in aspect of awareness and mentality, it intensively needs to organize the course contents intelligent and present them to the learner well.

The aim of intelligent E-learning is to realize the customized and adaptive E-learning using of course content, learner type and education method [6]. This system can recognize the student type. Then it can chooses appropriate course content from knowledge base and present the contents in proper style to the learners. Figure 1 shows the process of ITS.

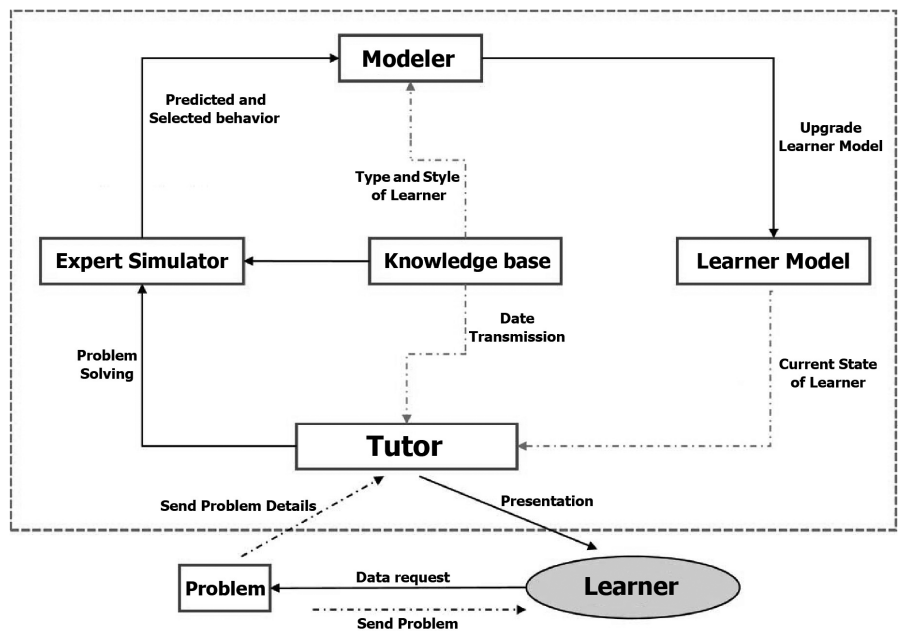

**Fig. 1. Process of Intelligent Tutoring System**

Some available intelligent E-learning systems are introduced in the following to handle the pragmatics of three elements: content, learner model and education methods for adaptive and customized learning.

- **VCA System** [7]: To train the learners, it considers individual differences and talents of learners to develop a virtual classmate agent.
- **SQL-Tutor System** [8]: It has been developed by guided exploration. This system selects some questions in basis of learner's model. Then it evaluates learner's answer. It updates the model based on answer validity. Choosing questions would be repeated based on model.
- **Lisp-Tutor system** [9]: This system guides the learner intelligently in each step of problem solving without considering his answers. This system tries to teach LISP.
- **DeSIGN System** [10,11]: This is a software to teach American Language to deaf learners. This system teaches English words by elements of "train-test" and "teacher" graphically. This system is used in Pittsburg deaf School now.
- **EIAS system** [12]: It is as adviser for collaborative learning.
- **CAES system** [13]: It has been developed by integration of shipping simulation and intelligent decision system. Its task is to teach shipping to captains in virtual turbulent sea.
- **UC-Links** [14]: It is an intelligent system to present the courses in the universities.
- **GENITOR system** [15]: It is the generator of training programs.
- **ICATS** [16]: This system coordinates the expert system with multimedia system in an intelligent learning system.

## 3. Adaptive E-learning

The adaptive E-learning is very important in order to improve efficiency and effectiveness of educational environments. These systems can also be responsible and compatible with the heterogeneous population of learners. An E-learning which is efficient, adaptive and dynamic can recognize learning style of learner by pedagogy principles. It can adapt the learners with current status of system. Then it changes its behavior dynamically and presents the learning concepts according to learner's model. This way leads to improve learning rate finally.

Some psychologist and pedagogy researchers applied many models in adaptive E-learning systems to model behavior and learning style of learners. This model has many advantages in comparison with others. They are proper analysis, recognition of ideal learning style and application of educational science in modeling [17]. Next section introduces styles of learning based on Jakson model and questionnaire.

## 3.1 Learning styles

An adaptive E-learning system is based on accurate recognition of behaviors and individual characteristics. In addition of aptitude, personality and behavior, learning style is very important as well [17].

This paper is based on five learning style which is summarizes in Table 1:

- **Sensation seeking (SS):** These people are impulsive and aren't patient. New situations are exciting so that they can't wait and would like to experience and explore it immediately. They believe to action and perform multiple tasks simultaneously. These people would rather to explore their environment by themselves and also learn by test and error.
- **Goal Oriented achievers (GOA):** They adjust certain and difficult goals. They try to increase their abilities by attaining skills and collecting required cognitive resources to realize their goals. They think that troubles are as instructive challenges. Furthermore they believe that can realize to whatever they want.
- **Emotionally Intelligent Achievers (EIA):** Emotional independence and rational thinking are prominent characteristic of them. They are patient learners who have the best efficiency after perceiving of logic behind a problem. They can generalize well from one problem to others. They often divide a problem to smaller and intelligible ones in this process.
- **Conscientious Achievers (CA):** They are responsible and wise people. They can learn well by collecting, analysis and review some information before action. They prefer to analyze all problem aspects. Thus they can relate discrete data to each other and avoid making a mistake. These people usually have extensive knowledge in areas of interest.
- **Deep learning Achievers (DLA):** they have deep perception of concepts. They want to know how can use previous taught practically. They may learn well when would be aware of note value. So they can test that theory or idea. In fact, learning is difficult for them, when they don't know the target [17].

**Table 1. Summarizes the Mentioned Learning Styles**

| Learning style | Comment |
| --- | --- |
| Sensation Seeking | They believe that experiences create learning. |
| Goal-oriented Achievers | They set difficult and certain target. They have self-confidence to achieve them. |
| Emotionally Intelligent Achievers | They are rational and goal-oriented instead of dependent and sentimental. |
| Conscientious Achievers | They are responsible and insight creator. |
| Deep learning Achievers | They are interested in learning highly. |

## 4. Proposed system

A web-based, adaptive and intelligent tutor is an E-learning system based on web which can be used remotely and ETEW. It can determine learner type (especially in aspect of learning style), learning content and presentation technique adaptively. So that it be updated automatically with learner's characteristics and behavior. This system uses traditional intelligent E-learning. The first E-learning system which is web-based and intelligent, has been reported on 1995[5,18]. Learning all courses is customized well at home via web in this system. So learners can solve some examples and proper exercises ETEW. Finally he can attend in course examination which would be virtual or physical. Figure 2 shows the elements of intelligent tutor.

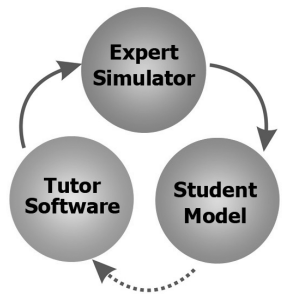

**Fig. 2. Elements of Intelligent Tutor**

### *4.1 Learning environment*

Learner can visits website of intelligent virtual school by authentication and logging in the system. An intelligent Graphical User Interface (GUI) is an interface between learners and intelligent tutor. This section of system can affect learning efficiency. An intelligent virtual class has some properties like graphical properties, audio and video to make learning attractive. Moreover, some tools are available to simplify learning process. Learners can communicate well with this inanimate and non-physical system by these tools. Some facilities are:
– Computer games- preferably intellectual games and commensurate with the level of learner
– Frequently asked questions (FAQ)- which consists of commonly questions and proper answers
– Video chat and email- for visual communication between tutor and learners

## *4.2 Training method*

Knowledge of expert tutor includes of two parts, course knowledge (learning content) and learning technique. Course knowledge is theoretic information, technical content and probably experiments which expert tutor notes. Learning technique is some experiences which he have got during teaching years [19]. An expert tutor determines learner level in according to IQ, understanding, behaviors, talent and individual characteristic like physical class. Learner level consists of "*weak*", "*slow learner*", "*smart*", "*genius*" and so on. Tutor teaches educational content corresponding to learner level in proper method such as film, dynamic view, and game and even bringing up puzzle while he get feedback from learner during training. So learner level may be changed. Tutor helps learner to learn by "*the best way*" in proposed system.

The expert tutor offers an education method based on learner's type. Furthermore each course section has individual significance which is different with the others. Tutor often determine different scores for variant sections according to education method. Moreover he marks highest score to the most important section of course in all education methods.

There are two approaches in E-learning development. In the first one, a problem comes and some examples would be solved then. Finally the learners try an exam. In second approach, content is divided to some parts such as chapter, section, important subsection and so on. The learners take an exam at the end of learning. It is clear the second approach is more effective and has higher level than the first one. In this paper, the proposed system uses the second approach. The smallest part of any topic which can't divide more is called "concept". It is usually equivalent a lesson in physical class. Educational concepts transfer to knowledge base in this system. Then the system can distinguish all concepts and relocate all parts. Sometimes a lesson is needed to repeat, relocate or even remove for a learner. Most of available systems guide a learner to a special aim intelligently in learning process. While only a few intelligent systems provides selecting subsections of a concept for a learner.

This system uses a three layered structure to offer and implement a concept:

1. Pre-test
2. Learning concept
3. Post-test

The pre-test includes of some questions planned by an expert tutor to determine learner's primary knowledge level. The learning concept depends on learner level. So the best method to train a learner is determined. Then learning process starts up. After learning is done, a post-test evaluate the learner by some questions. Figure 3 shows block diagram of proposed system.

## *4.3 Learning styles*

Learner evaluation is significant. It has two levels, conceptual and objective. Evaluating in concept level refers to learner understanding of lesson concept and evaluating in objective level denotes to learner understanding of lesson topic. Knowledge level of learner is determined with concept level and objective level. The tutor can extract proper questions from question base through an expert system, pre-test and post-test. He notes that a specific score is given to each question.

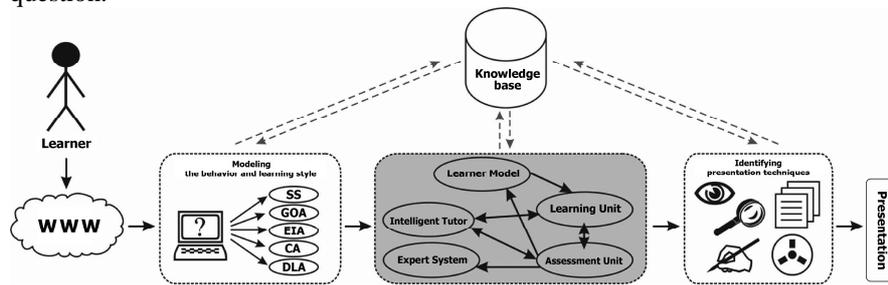

**Fig. 3. Block Diagram of Proposed System**

Selecting question should satisfy some rules. First, the questions should not be repetitious even if a learner would be trained one concept several times. Second, the question must be planned for all sections of a concept entirely. Third, expert tutor plans questions in all level. Sequence, number and level of questions are determined according to learner level and learning type intelligently. Sum of scores is calculated and learner level is determined after answering the questions.

Table 2 presents five categories of learner's knowledge level about a concept [20, 21]:

**Table 2. Categories of Knowledge Level**

| Knowledge Level | Score |
|---|---|
| Excellent | 86-100 |
| Very good | 71-85 |
| Good | 51-70 |
| Average | 31-50 |

This system updates the learner's model during progress of question answering. This system can also save last academic status of learner and all his learning records.

## 5. Conclusion

A web-based, adaptive and intelligent tutor by expert system was presented in this research. Previous E-learning systems offer predefined and static learning concept sequentially to learners. While proposed system can adapts with learning styles (i.e. Sensation Seeking, Goal Oriented achievers, Emotionally Intelligent Achievers and Conscientious Achievers), aptitude, characteristics and behaviors of a learner. It acts as an intelligent tutor which can perform three processes - *pre-test*, *learning concept* and *post-test* - according to characteristic learner. This system uses expert simulator and its knowledge base as well. It is also web-based which leads to be simple learning, low-cost, available everywhere and every time. Consequently thousands of students can learn simultaneous and integrated efficiently.

Nowadays the most educational systems try to be electronic, online, intelligent, adaptive and dynamic. The proposed system tries to get these properties. Moreover it doesn't have any drawback of previous system and human expert tutor. It can improve efficiency of pedagogy and education too. In other words, it helps learners to study in "*the best way*".